\documentclass[12pt]{iopart}
\usepackage{iopams}
\usepackage{cite}
\usepackage{graphicx}
\usepackage[outdir=./]{epstopdf}
\graphicspath{{figure/}}
\usepackage{subfigure}
\usepackage{makecell}
\usepackage{threeparttable}
\usepackage{siunitx}
\usepackage{hyperref}
\usepackage{xcolor}
\usepackage{ulem}
\usepackage{soul}
\definecolor{bluee}{rgb}{0.062,0.086,0.796}
\hypersetup{colorlinks=true,linkcolor=bluee,urlcolor=bluee,anchorcolor=bluee,citecolor=bluee,CJKbookmarks=True}

\newcommand{\R}[1]{\textcolor{red}{#1}}

\soulregister{\cite}{1}
\newcommand{\D}[1]{\textcolor{red}{\sout{#1}}}

\renewcommand{\D}[1]{}
\renewcommand{\R}[1]{#1}

\begin{document}
\title[Experiment demonstration of tilt-to-length coupling suppression by beam-alignment-mechanism]{Experiment demonstration of tilt-to-length coupling suppression by beam-alignment-mechanism}
\author{Peng Qiu$^{1}$, Xiang Lin$^{1}$, Yurong Liang$^{1}$, Hao Yan$^{1,*}$, Haixing Miao$^{2}$, Zebing Zhou$^{1}$}
\address{$^1$ MOE Key Laboratory of Fundamental Physical Quantities Measurement and Hubei Key Laboratory of Gravitation and Quantum Physics, PGMF and School of Physics, Huazhong University of Science and Technology, Wuhan 430074, China.}
\address{$^2$ State Key Laboratory of Low Dimensional Quantum Physics, Department of Physics, Tsinghua University, Beijing, China}
\ead{yanhao2022@hust.edu.cn}
\begin{abstract}
Tilt-to-length (TTL) noise, caused by angular jitter and misalignment, is a major noise source in the inter-satellite interferometer for gravitational wave detection. However, the required level of axis alignment of the optical components is beyond the current state of the art. A set of optical parallel plates, called beam alignment mechanism (BAM), is proposed by LISA to compensate for the alignment error. In this paper, we show a prototype design of the BAM and demonstrate its performance in a ground-based optical system. We derive the BAM theoretical model, which agrees well with the numerical simulation. Experimental results reveal that the BAM can achieve lateral displacement compensation of the optical axis with a resolution of \SI{1}{\micro\meter} across a \D{dynamic} range of about \SI{0.5}{\milli\meter}. Furthermore, the TTL coefficient is reduced from about \SI{0.3}{\milli\meter/\radian} to about \SI{5}{\micro\meter/\radian}, satisfying the preliminary requirements for LISA and TianQin. These findings confirm the efficacy of the BAM in suppressing TTL noise, offering a promising solution for space-based gravitational wave detection.
\end{abstract}

\submitto{\CQG}

\section{Introduction}
Gravitational wave detection has opened new windows into the universe, enabling the observation of phenomena such as black hole mergers and neutron star collisions \cite{ligo_scientific_collaboration_gwtc-3_2023}. Space-based gravitational wave detectors like LISA \cite{colpi_lisa_2023}, TianQin \cite{luo_tianqin_2016, mei_tianqin_2021} and Taiji \cite{hu_taiji_2017} are poised to extend the observation window to low-frequency gravitational waves, enabling the detection of massive black hole mergers and the early universe's stochastic gravitational wave background. However, despite the potential breakthroughs, space-based GW detectors face numerous technical challenges, with tilt-to-length (TTL) coupling noise emerging as one of the primary limitations for their sensitivity.

TTL coupling arises when angular misalignments in the optical system introduce unwanted length changes in the interferometric measurements \cite{hartig_geometric_2022, hartig_non-geometric_2023}. \R{Among the various TTL coupling mechanisms, the first-order term in the Taylor expansion, arising from the alignment of the optical axes of different modules (such as the test mass, optical bench, telescope, etc.), is one of the most dominant\cite{paczkowski_postprocessing_2022}.}
\D{In current research, TTL coupling is categorized into geometric and non-geometric TTL. Among the various coupling mechanisms, the first-order TTL coupling, often referred to as the ‘piston effect’, is the most significant. This effect occurs when angular tilts between different optical modules result in apparent path length changes due to the non-perfect overlap of laser beams. The primary contributors to this coupling include the alignment of the telescope, optical bench, and test masses in the interferometric system. These misalignments can introduce length noise that dominates over the desired gravitational wave signals.} To mitigate this issue, the LISA team has proposed the use of a dedicated beam alignment mechanism (BAM) to \D{actively} adjust the optical alignment and reduce TTL coupling \cite{brzozowski_lisa_2022,colpi_lisa_2023}. The BAM was designed to suppress the TTL coupling coefficient of the scientific interferometer from $\SI{8.5}{\milli\meter/\radian}$ to $\SI{2.3}{\milli\meter/\radian}$\D{\mbox{\cite{colpi_lisa_2023}}}\R{\cite{brzozowski_lisa_2022}}. In TianQin mission, the interstellar displacement sensitivity requirement is higher, the TTL suppression requirement is higher, and the technical challenge is greater \cite{gong_concepts_2021}. \R{Based on the budget for the TTL coupling noise in TianQin mission, we initially set the compensation range at \SI{0.5}{\milli\meter}, with a compensation precision of \SI{1}{\micro\meter}.}

Currently, the BAM is in its conceptual design phase, and comprehensive theoretical analyses or experimental validations of its performance are still lacking. The most significant challenge in implementing BAM lies in achieving precise two-dimensional lateral position adjustment of an Optical axis while maintaining its constant optical path length. This necessitates the ability to make millimeter-scale adjustments with micron-level precision, a requirement that pushes the limits of current optical and mechanical design. Furthermore, the mechanism must operate in the harsh environment of space, where thermal fluctuations and mechanical stresses add further complexity to maintaining alignment.

In this study, we investigate the application of a BAM as a potential solution for mitigating first-order TTL noise. The BAM employs parallel plates to dynamically adjust and correct the beam path. By fine-tuning these components, we aim to minimize transverse displacement and thereby reduce first-order TTL coupling.

The paper is organized as follows. \Sref{sec_theory model} provides a theoretical analysis of effects associated with BAM. Numerical simulations were also performed to calculate the corresponding outcomes. \Sref{sec_beam alignment tests}  details experiments involving CCD\R{(Charge-Coupled Device)} testing to investigate the compensation range of BAM. \Sref{sec_ttl tests} describes the design of a dual-beam interferometry apparatus to validate BAM's suppression effect on first-order TTL coupling. Finally, \Sref{sec_conclusion} contains the conclusions and discussion.

\section{Theory model}
\label{sec_theory model}
In the space gravitational wave interferometer system, the BAM employs a dual glass plate design to maintain a constant optical path length. A laser beam passes obliquely through the surface parallel plates, and by controlling the rotation angle of the two plates, the lateral position of the optical axis is adjusted. Additionally, to enable arbitrary adjustments of the optical axis position within a two-dimensional plane, each plate is designed to operate independently.

\subsection{Ideal model}
\begin{figure}[ht!]    \centering\includegraphics[width=0.85\textwidth]{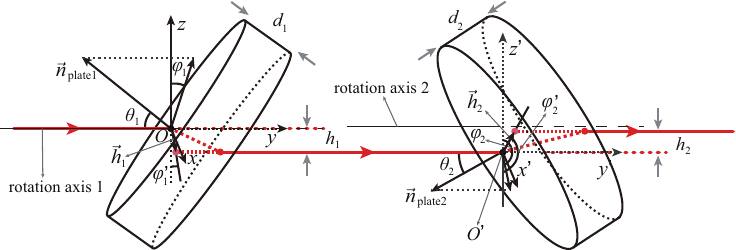}
    \caption{\R{Optical schematic of the BAM, consisting of two independently inclined plates that adjust the laser beam’s lateral position.}}
    \label{fig1}
\end{figure}

The schematic of the BAM is shown in \fref{fig1}. The target laser beam passes through two surface parallel plates at an oblique angle, resulting in a lateral displacement of the optical axis. 
In this simplified model, the glass surfaces are parallel, and the rotation axis aligns with the incident optical axis. Therefore, the exiting optical axis is parallel and symmetric to the incident optical axis.
For the first plate, the incident beam has thickness $d_1$ and refractive index $n_1$ at an incidence angle of $\theta_1$. The lateral displacement of the optical axis, $\vec{h}_1$, is given by:
\begin{equation}
    h_1 = d_1 \sin \theta_1 \left(1 - \frac{\cos \theta_1}{\sqrt{{n_1}^2 - \sin^2 \theta_1}} \right).
    \label{eq_dh}
\end{equation}

\begin{figure}[ht!]    \centering\includegraphics[width=0.5\textwidth]{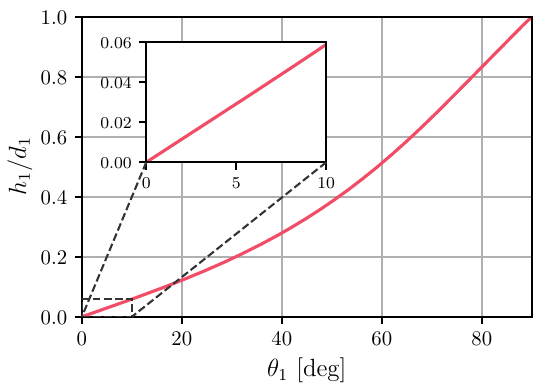}
    \caption{Relationship between the lateral displacement \D{$h$} \R{$h_1$}and the incidence angle \D{$\theta$} \R{$\theta_1$}.}
    \label{fig2}
\end{figure}

\Fref{fig2} illustrates the relationship between lateral displacement and incidence angle $\theta$. In the range of \SIrange{0}{10}{\degree}, lateral displacement increases approximately linearly. By adjusting the thickness of the plate and the angle of incidence, the BAM’s compensation range can be finely controlled. This allows for precise lateral adjustments of the optical axis, essential for space-based gravitational wave detection missions.

\begin{figure}[ht!]
    \centering
    \includegraphics[width=0.45\textwidth]{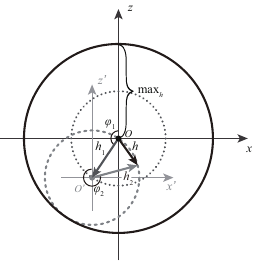}
    \caption{Compensation coverage of the BAM. $\vec{h}_1$ represents the compensation position of the first plate, and $O'$ is the endpoint of $\vec{h}_1$. This vector can rotate as an angle $\varphi_1$ within the $x\text{-}O\text{-}z$ plane around point $O$ as the plate rotates, forming a compensation range that is a circle with center $O$ and radius $h_1$. Similarly, $\vec{h}_2$ is the compensation position of the second plate, forming a circular compensation range with center $O'$ and radius $h_2$. When the two plates are combined, the overall compensation position is denoted as $\vec{h}$. Since both plates can rotate independently, the final compensation range forms an annulus centered at $O$, with radii ranging from $|h_1 - h_2|$ to $|h_1 + h_2|$.}
    \label{fig3}
\end{figure}

\R{In the $x\text{-}O\text{-}z$ plane, as shown in \fref{fig1} and \ref{fig3}, both the compensation direction, defined as the angle $\varphi'_1$ between the refracted beam and the -$z$ axis, and the rotation angle $\varphi$ of plate, which is the angle between the normal vector of the plate surface and the $z$ axis, are projected within this plane.}

As the plate rotates around the optical axis, the lateral displacement amplitude remains constant, while its direction rotates by the same angle, satisfying:
\begin{equation}
    \varphi'_1 = \varphi_1.
    \label{eq_varphi}
\end{equation}
As a result, the position of the exiting optical axis traces a circular trajectory of radius \R{$h_1$} around the incident optical axis. In space-based gravitational wave detection, the optical-axis compensation must be adjusted in two lateral degrees of freedom. Thus, we need a second plate.

In a new coordinate system $x'\text{-}O\text{-}z'$, similar to the first plate, the lateral displacement of the optical axis and compensate angle are given by:
\begin{equation}
    h_2 = d_2 \sin \theta_2 \left(1 - \frac{\cos \theta_2}{\sqrt{{n_2}^2 - \sin^2 \theta_2}} \right),
\end{equation}
\begin{equation}
    \varphi'_2 = \varphi_2.
\end{equation}
The total lateral displacement of the optical axis $h$ is the sum of the lateral displacements produced by the dual plates, given by:
\begin{equation}
    \vec{h} = \vec{h}_1 + \vec{h}_2.
    \label{eq3}
\end{equation}
By adjusting the parameters $d_1$, $d_2$, $n_1$, $n_2$, $\theta_1$ and $\theta_2$ to ensure $h_1=h_2$, the position of the exiting optical axis of the BAM resides within a circle centered at point $O$ with a radius of $2h_1$, as shown in \fref{fig3}. Assuming ideal conditions with $d_1=d_2=$\SI{3}{\milli\meter}, $n_1=n_2=$\SI{1.45}{}, and $\theta_1 = \theta_2=$\SI{8.5}{\degree}, the resulting compensation radius \R{$h$ is about \SI{ 280}{\micro\meter}.}

\newpage
\subsection{Error analysis}
\D{Two key factors affect the practical adjustment of the BAM's optical axis compensation. First, if the rotation axis is not perfectly aligned with the incident optical axis, the compensation of the outgoing beam will vary. Second, any deviation from parallelism between the two surfaces of the plate leads to a misalignment between the incident and exiting beams. In this analysis, we assume that the plate surfaces are sufficiently flat to minimize additional distortion.}
\R{Space-based gravitational wave detection requires BAM to achieve sub-millimeter compensation and alignment precision at the micron level for position and the micro-radian level for pointing. Practical error sources must be considered. Two main non-negligible factors affect BAM optical axis compensation. First, misalignment between the rotation axis and the incident optical axis alters the compensated outgoing beam. Second, non-parallel plate surfaces misalign the incident and exiting beams. The following sections provide a detailed modeling and analysis of these errors. For simplicity, we assume sufficiently flat glass surfaces to minimize wavefront distortions.}

\subsubsection{Axis misalignment}
\label{sec_axis}
\begin{figure}[ht!]   \centering\includegraphics[width=0.56\textwidth]{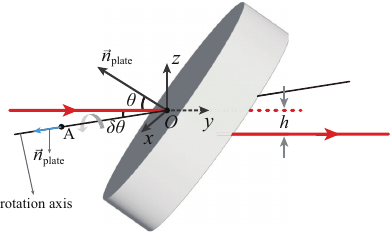}
    \caption{Axis Misalignment: The plate's rotation axis is misaligned by an angle $\delta\theta$ relative to the beam axis.}
    \label{fig4}
\end{figure}

In cases where the rotation axis is misaligned by an angle $\delta \theta$ relative to the incident optical axis, as shown in \fref{fig4}, the compensation behavior of the beam is altered. A coordinate system is defined with the point of incidence $O$ as the origin, and the incident beam propagating along the $y$ axis. 

\R{When the plate pane rotates around the axis of rotation, the angle of incidence will change accordingly:
\begin{eqnarray}
    \theta(\varphi) = \arccos n'_y
\end{eqnarray}
More math detail can be find in \ref{app_axis}. }Since the surfaces of the plate remain parallel, $\theta(\varphi)$ affects both the magnitude and direction of the optical axis compensation as follows:
\begin{equation}
    \cases{
    h'(\varphi)=d \sin \theta \left(1 - \frac{\cos \theta}{\sqrt{{n}^2 - \sin^2 \theta}} \right)\\
    \varphi' = \arctan \frac{n'_x}{n'_z}}
    \label{eq_error2}
\end{equation}

\begin{figure}[ht!]
    \hspace{3cm}
    \includegraphics[width=0.65\textwidth]{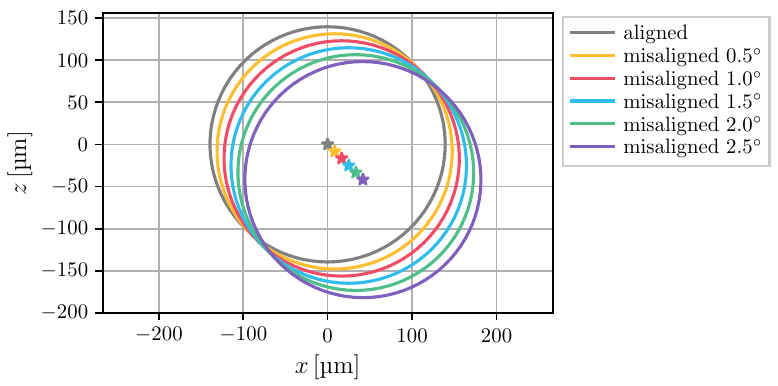}
    \caption{Simulation of the compensation range with axis misalignment. Asterisks indicate the fitted centers of the corresponding compensation ranges.}
    \label{fig5}
\end{figure}

\Fref{fig5} show the simulation results of the compensation angle and radius fluctuate as the rotation angle $\varphi$ increases, with the magnitude of these fluctuations growing as the misalignment angle increases. For small misalignment angles, the compensation range remains nearly circular. However, as the misalignment angles increase, the center of the circular compensation range shifts along the $z$-axis and $x$-axis, respectively.\R{This effect causes lateral shifts and compression of the compensation range, which results in beam misalignment. Linear regression shows a relationship between the compensation centroid offset $\Delta x_c$ and the axis-misalignment angle $\delta\theta$, with a slope of $\SI{967.29}{\micro\meter/\radian}$. Compression, measured by the radial deviation $r$ from the fitted center, fluctuates by $\approx\SI{0.05}{\micro\meter}$ at $\delta\theta = \SI{1}{\degree}$, making it negligible compared to lateral displacement under these conditions. Given the specified compensation accuracy of \SI{1}{\micro\meter}, our analysis suggests  that the critical axis-misalignment threshold should be constrained to $\delta\theta < \SI{1}{\milli\radian}$ to ensure optimal beam alignment.}

\newpage
\subsubsection{Surface parallelism error}
\label{sec_para}
\begin{figure}[ht!]    
    \centering\includegraphics[width=0.52\textwidth]{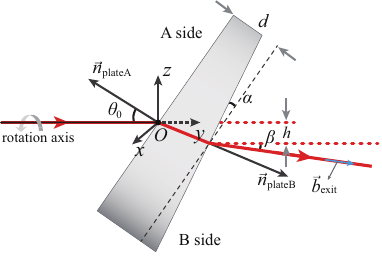}
    \caption{The surface parallelism error caused by the plate’s wedge angle $\alpha$, leading to a beam-axis tilt $\beta$.}
    \label{fig6}
\end{figure}

The second factor affecting beam alignment is the non-parallelism of the plate surfaces, as illustrated in \fref{fig6}. In this scenario, the plate thickness is defined as the distance between the two surfaces. The parallelism error, denoted as $\alpha$, represents the angle between the surfaces, which introduces a misalignment between the incident and exiting beams, denoted by $\beta$.

\R{When the plate rotates, the angle of incidence $\theta_0$ is maintained, signifying that the compensation radius remains unchanged. At the same time, the constant angle between the normal vector of the plate, $\vec{n}_{\text{plateB}}$, and the y-axis results in a stable beam deflection angle $\beta$ induced by the plate. }

\R{With the mathematical detail in \ref{app_surf} , }the compensation value and angle are calculated as
\begin{equation}
    \cases{
    h' = d \sin \theta \left(1 - \frac{\cos \theta_0}{\sqrt{{n}^2 - \sin^2 \theta_0}} \right)\\
    \varphi' = \varphi\\
    \vec{b}_\text{exit} = \mathbf{M}(\vec{n'}_{\rm plateB}) \cdot \mathbf{M}(\vec{n'}_{\rm plateA}) \cdot \vec{y}
    } ,
\end{equation}
\R{where $\vec{b}_\text{exit}$ is the direction vector of the exiting beam. And beam deflection angle $\beta$ can be express as}
\begin{eqnarray}
    \beta &=& \arcsin\left( \sin\theta_0 \cos\alpha + \sqrt{n^2 - \sin^2\theta_0} \cdot \sin\alpha \right) - \theta_0 - \alpha \nonumber \\
    &\approx&\left(\frac{\sqrt{n^2 - \sin^2\theta_0}}{\cos\theta_0} -1 \right) \alpha
\end{eqnarray}

The surface parallelism error, as illustrated in \fref{fig6}, arises from the wedge angle of the plate. When the plate is fixed, the wedge angle is characterized by the normal vector of surface B, represented as $\vec{n}_{\text{plateB}}$. The projection of $\vec{n}_{\text{plateB}}$ onto the $x\text{-}O\text{-}z$ plane can vary in any direction. Therefore, the surface parallelism error depends on both the wedge angle's magnitude and the orientation of $\vec{n}_{\text{plateB}}$. The wedge angle determines the amount of tilt introduced to the optical axis, while the orientation of $\vec{n}_{\text{plateB}}$ defines the direction of the tilt. In the simulation, the wedge angle is set to $\pm \SI{30}{\arcsecond}$, and $\vec{n}_{\text{plateB}}$ is positioned within the $x\text{-}O\text{-}z$ plane.

\begin{figure}[ht!]
    \hspace{1.2cm}
    \includegraphics[width=0.9\textwidth]{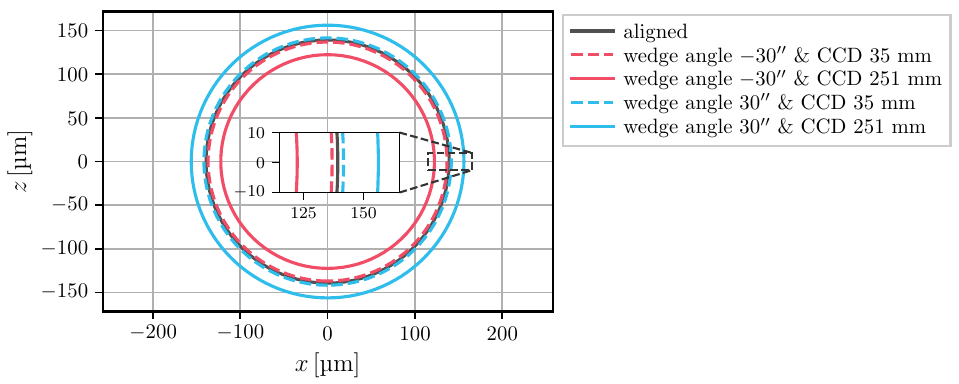}
    \caption{Simulation of the compensation range with surface parallelism error. solid lines and dashed lines represent opposite directions of the plate wedge angle.}
    \label{fig7}
\end{figure}

\R{The simulation results in \fref{fig7} show that surface parallelism errors of the plate slices induce an additional tilt in the outgoing beam axis, with the tilt direction determined by the orientation of $\vec{n}_{\text{plateB}}$. In space-based gravitational wave detection, the BAM is primarily used to compensate for pupil misalignments between the telescope and the optical interferometer bench. Any additional angular deviation in the outgoing beam axis reduces the interference contrast and increases TTL coupling errors. Given the TianQin specifications, the additional angular offset caused by the BAM should not exceed \SI{30}{\micro\radian}.}


\section{Beam alignment tests}
\label{sec_beam alignment tests}
\subsection{Optical setup}
\begin{figure}[ht!]
    \centering
    \includegraphics[width=0.7\textwidth]{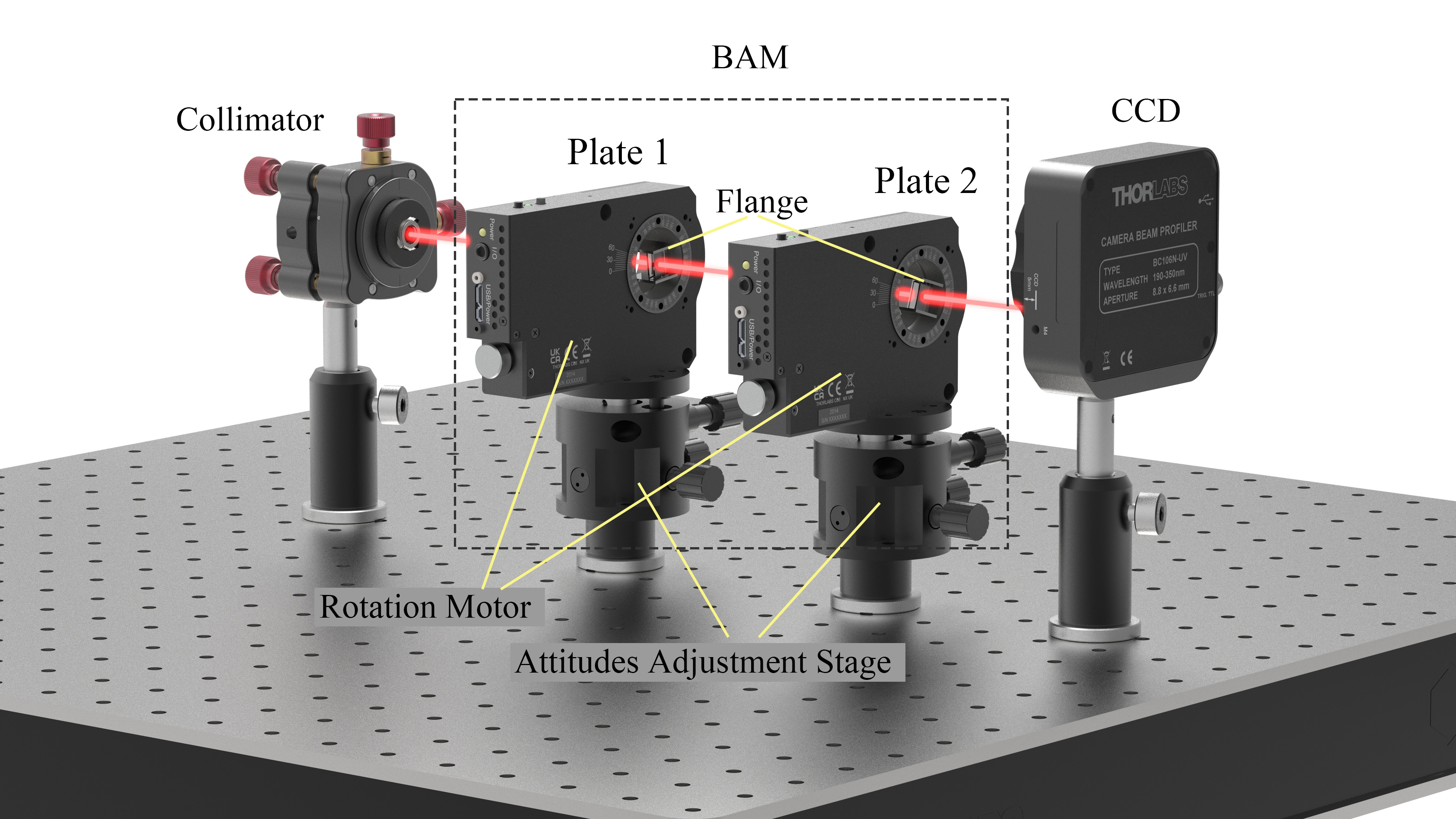} 
    \caption{Schematic diagram of the BAM for experimental validation. The Gauss beam, originating from the collimator on the left, sequentially traverses through dual plates positioned on a rotation motor, ultimately being detected by the CCD\R{Charge-Coupled Device}. The rotational axis is adjusted using an \D{tip, tilt, and rotation stage (not shown)}\R{attitudes adjustment stage } beneath the rotation \D{stage} \R{motor} to align as closely as possible with the beam axis. When the two \D{rotational mountings} \R{rotation motor} move independently, the beam experiences lateral displacement within a plane constrained by a \D{circular} \R{annulus} range.}
    \label{fig8}
\end{figure}

The optical setup for beam alignment tests has been meticulously designed to evaluate the performance and precision of the BAM. Schematic diagram of the BAM experiment is shown in \fref{fig8}. Central to this experiment are two identical K9 plates measuring \SI{10}{\milli\meter} $\times$ \SI{10}{\milli\meter} $\times$ \SI{3}{\milli\meter}. These plates are mounted at an angle on high-precision motorized rotation motor, which provide angular control with a resolution of about $\SI{0.05}{\degree}$. This level of precision enables fine adjustments to the beam path, essential for achieving accurate alignment.

\begin{figure}[ht!]
    \centering
    \includegraphics[width=0.6\textwidth]{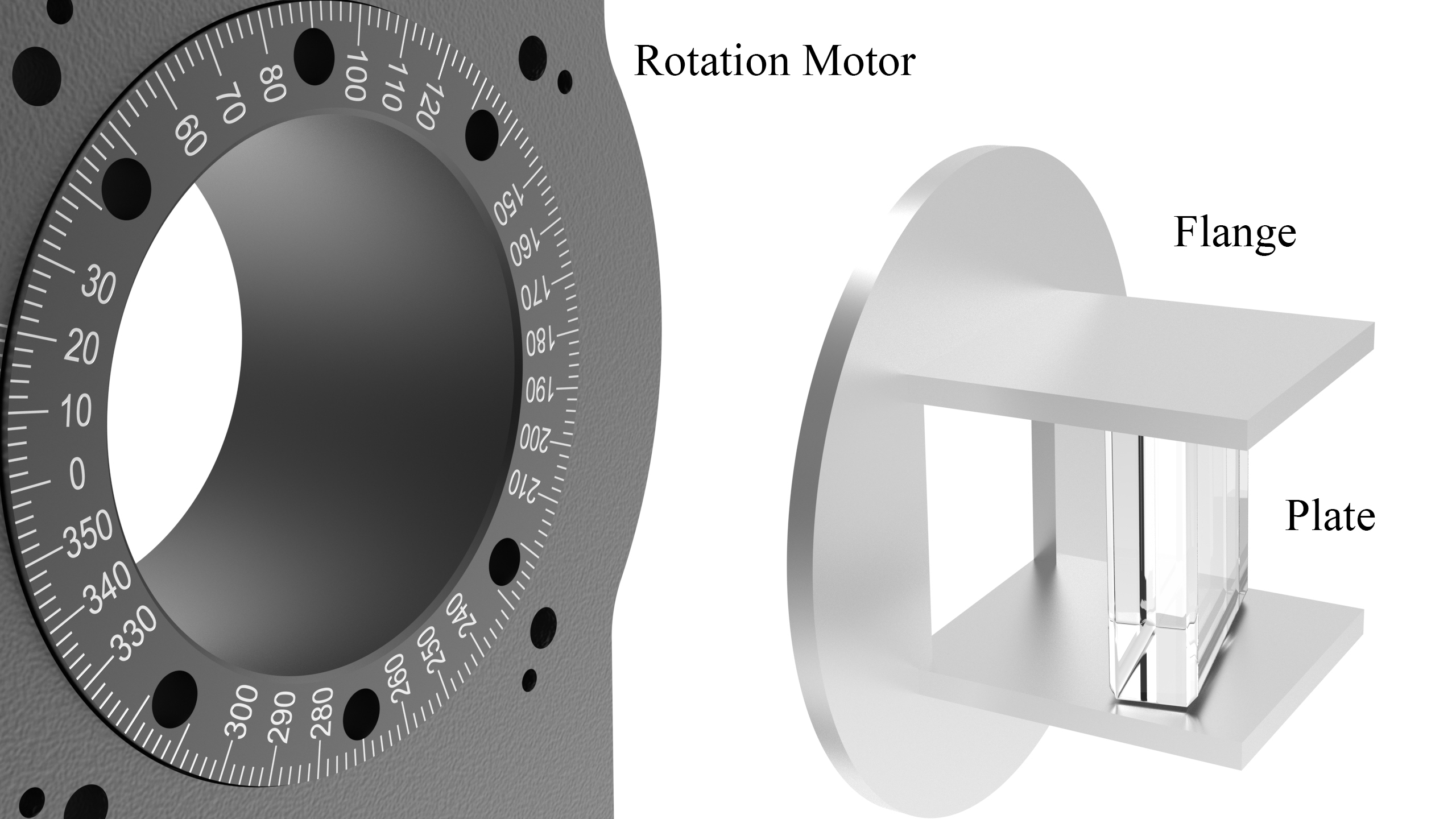} 
    \caption{Schematic diagram of the flange connecting the plates to the rotation motor. The flange is secured to the rotation motor using retaining rings. Initially, the plates are clamped by the inherent stress of the flange and are later bonded to the flange with epoxy resin after adjusting the incident angle.}
    \label{fig9}
\end{figure}

Initially, the plates are set at an approximate angle of \SI{8.5}{\degree} relative to the incident beam, facilitating effective control of beam displacement during rotation. \R{Since the thickness of each plate cannot be perfectly uniform, it is necessary to adjust the incident angle to ensure that the compensation radii of the two plates become approximately equal. We designed the flange shown in \fref{fig9} to connect the plates to the rotation motor; the flange is secured to the rotation motor using retaining rings. Initially, the plates are clamped by the inherent stress of the flange, and after adjusting the incident angle, they are bonded to the flange using epoxy resin.}

\R{To adjust the compensation radii, it is necessary to align the rotation axis with the incident light axis. First, we adjust the beam to be as parallel as possible to the optical table and to be vertically incident on the CCD, using the CCD to detect the center position of the beam as a reference point. Next, we place a plate mounted on the rotation motor between the CCD and the collimator, causing the center of the beam spot on the CCD to shift accordingly. According to our simulation results, if the rotation axis and the incident light axis are not aligned, the final compensation range will be offset as a whole. This offset manifests as a displacement between the center of the compensation range and the reference point.}

\R{Therefore, we rotate the plate through a full revolution, collect the corresponding beam center positions on the CCD, and calculate the center of the compensation range by fitting these positions. Using the attitude adjustment stage, we adjust the rotation axis of the rotation motor until the fitted center coincides with the reference point. This alignment indicates that the rotation axis and the incident light axis are now properly aligned.}

\R{Subsequently, the relative angle between the plate and the flange can be adjusted to fine-tune the compensation radius. After each adjustment, the plate is rotated once, and the compensation radius is determined by fitting the data from the CCD.}

\R{However, to achieve the precise vertical incidence of the beam on the CCD, we manually adjust the CCD position. This ensures the beam is as perpendicular as possible to the surface of the CCD. Once this adjustment is made, the plate is rotated through a full revolution, and we collect the beam center data corresponding to each position on the CCD. To account for any distortion due to the rotation, we apply a projection transformation to the data. This transformation corrects the data to be as circular as possible, and the projection coefficient corresponding to the CCD position at this time is calculated. All subsequent data processing will then be adjusted by multiplying the projection coefficient, ensuring accurate results that account for any deviations in the CCD's orientation.}

\R{To achieve the precise vertical incidence of the beam on the CCD, we manually adjusted the CCD position to approach perpendicular alignment (though perfect verticality cannot be guaranteed). Following this coarse alignment, the plate was rotated through a complete revolution while recording centroid positions of the beam center on the CCD. }

\R{Due to beam path rotation about the plate's axis and potential residual CCD tilt, the collected coordinates formed elliptical distributions rather than perfect circles. We therefore applied an affine transformation to rectificate the coordinates:}

\begin{equation}
x' = k_{xx}x + k_{xz}y, \quad
z' = k_{yz}x + k_{zz}y
\end{equation}
\R{where the transformation coefficients $k_{ij}$ were determined through least-squares fitting to minimize deviations from circular symmetry. These coefficients inherently encoding the CCD's instantaneous spatial orientation were subsequently applied as multiplicative corrections to all experimental data during post-processing.}

The two BAM components are arranged sequentially in the optical path following a laser collimator, with a CCD positioned downstream of the second BAM component. This configuration allows for high-precision monitoring of the beam position, thereby enabling accurate measurement of beam displacement resulting from the combined effects of the BAM.

System control and data acquisition are managed by two specialized programs, which interface with the rotation mounts and CCD camera respectively. The software employs image processing algorithms to determine the centroid position of the beam spot on the CCD sensor, allowing for precise tracking of beam displacement relative to the pane rotation angles. The centroid position $(x_c, z_c)$ is calculated using the weighted average method:

\begin{equation}
    x_c = \frac{\sum_{i} \sum_{j} x_{ij} I_{ij}}{\sum_{i} \sum_{j} I_{ij}},
    \quad
    z_c = \frac{\sum_{i} \sum_{j} z_{ij} I_{ij}}{\sum_{i} \sum_{j} I_{ij}}
\end{equation}
where $x_{ij}$ and $z_{ij}$ are the coordinates of pixel $(i,j)$, and $I_{ij}$ is the intensity of that pixel. This method provides a robust and accurate determination of the beam position, accounting for variations in beam shape and intensity distribution.

\subsection{Test results}

To comprehensively assess the performance of the Beam Alignment Mechanism (BAM), we structured our experiments into three distinct parts: (1) evaluating the effects of axis misalignment using a single plate, (2) analyzing the impacts of surface parallelism using a single plate, and (3) investigating the combined effects of axis alignment with dual plates. Each experiment tracks the beam center across a full rotation of the plates, referred to as the compensation range.

In the initial experiments, we isolated the effects of individual components by removing one of the plates from the setup. This adjustment allowed for a clearer understanding of how each factor contributes to beam alignment.

\begin{figure}[ht!]
    \hspace{3cm} 
    \includegraphics[width=0.7\textwidth]{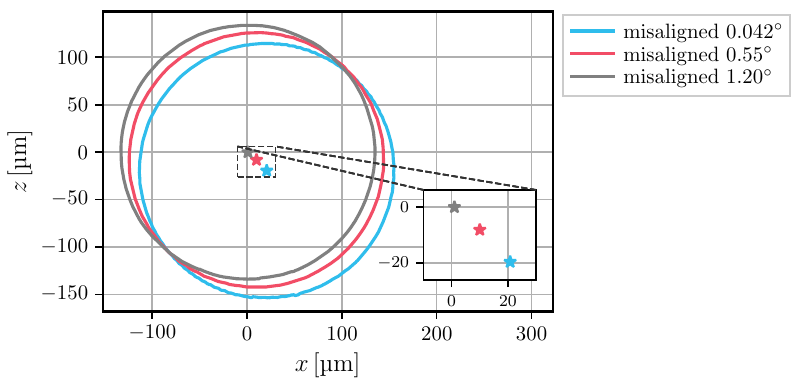}
    \caption{Compensation range with axis misalignment for a single plate. The gray line represents results with axis alignment, while the blue and red lines show results with both pitch and yaw angles of \SI{0.55}{\degree} and \SI{1.10}{\degree}, respectively. Asterisks indicate the fitted centers of the corresponding compensation ranges.}
    \label{fig10}
\end{figure}

The first experiment focused on the axis misalignment effect. We minimized the distance between the CCD camera and the plate to \SI{35}{\milli\meter} to \D{enhance measurement sensitivity} \R{mitigate effect caused by surface non-parallelism}. The results of this setup are shown in \fref{fig10}, \D{where the gray line represents the compensation range with axis alignment, while the blue and red lines indicate results with pitch and yaw misalignments of $0.55^{\circ}$ and $1.10^{\circ}$, respectively} \R{where the blue solid line represents the initial manual installation with lateral shifts in the compensation range of \SI{28.6}{\micro\meter}, whereas the red solid line demonstrates the improved alignment achieved through the attitude adjustment stage with a reduced offset of \SI{13.0}{\micro\meter}. The optimal alignment result (gray curve) exhibits a residual offset of merely \SI{1.0}{\micro\meter}, satisfying the system's precision requirements}. The fitted centers of the corresponding compensation ranges are denoted by asterisks.
\\

\R{Furthermore, angular misalignments of the attitude adjustment stage based on autocollimator calibration was about \SI{1.2}{\degree}, \SI{0.55}{\degree}, and \SI{0.042}{\degree} in the three tests, respectively. The linear correlation coefficient between lateral offset and angular misalignment was determined to be $\SI{975.7}{\micro\meter / \radian}$, in agreement with the theoretical model presented in \sref{sec_axis}.}
\begin{figure}[ht!]
    \hspace{3cm} 
    \includegraphics[width=0.76\textwidth]{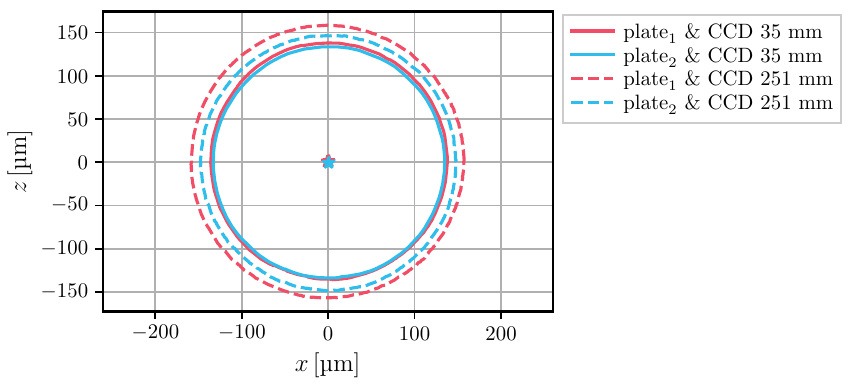}
    \caption{Compensation range with surface parallelism error for a single plate. Red and blue lines represent two different plates, with solid and dashed lines indicating distances of \SI{35}{\milli\meter} and \SI{251}{\milli\meter} between the CCD camera and the plate, respectively.}
    \label{fig11}
\end{figure}

Next, we examined the influence of surface parallelism error on beam alignment. \D{As depicted in \fref{fig11}, the} \R{The} rotation axis was pre-aligned to the optical axis, allowing us to analyze the tilt induced in the exiting beam due to misalignment between the front and rear surfaces of the plate. \D{We compared the compensation ranges at CCD distances of 35 mm and 251 mm, observing how this configuration affected beam behavior} \R{ The experimental procedure was as follows: First, the CCD was fixed in a set position so that the collimated laser beam from the collimator directly impinged on it, and the beam center was recorded. Next, the plate was inserted while ensuring that its distance from the CCD was maintained at \SI{35}{\milli\meter}. Using the method described in Section 3.1, the plate's rotation axis was aligned with the optical axis. The plate was then rotated a full 360° and the corresponding beam center on the CCD was recorded. Subsequently, the CCD was moved to a distance of \SI{251}{\milli\meter} from the plate, while the plate's rotation axis keeping aligned with the optical axis, and the plate was rotated one complete turn again with the beam center being recorded. Finally, with the CCD held fixed and the plate removed, the center position of the beam from the collimator was recorded}.

\R{\Fref{fig11} shows the compensation ranges for different plates and CCDs at distinct positions. According to the equation
\begin{equation}  
    \beta = \arcsin\left(\frac{\Delta r}{\Delta L}\right),  
\end{equation}
where $ \Delta L $ represents the distance between two CCD measurements and $\Delta r$ denotes the difference in compensation radii, the non-parallelism errors of the plate$_1$ and plate$_2$ panes are calculated to be $ \SI{96.4}{\micro\radian} $ and $ \SI{54.6}{\micro\radian} $, respectively. In subsequent experiments, we will select plate with smaller non-parallelism errors to conduct testing on the BAM prototype.}

\begin{figure}[ht!]
    \hspace{3cm}
    \includegraphics[width=0.7\textwidth]{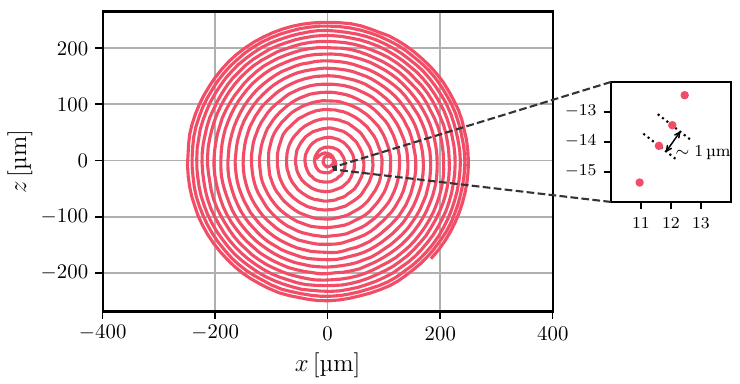}
    \caption{Compensation range with axis alignment for dual plates. The setup rotates two plates at different angular velocities over a period, creating a spiral-like pattern. Over time, the compensation range accumulates, forming a circular area with a diameter of about \SI{500}{\micro\meter}.}
    \label{fig12}
\end{figure}

In the third experiment, we explored the combined effects of axis alignment using two rotating plates. The setup, depicted in \fref{fig12}, was designed to create a spiral-like displacement pattern. As the two plates rotated at different angular velocities, the compensation range accumulated, forming an area constrained within a circular path. This design allowed us to visualize the cumulative impact of both plates on beam alignment.

\section{TTL tests}
\label{sec_ttl tests}
In previous sections, we have established the compensation range of the BAM structure for both single and dual plate configurations, particularly under the influence of axis and surface parallelism errors. This section delves into the application of BAM for inducing lateral displacement in interferometry to suppress first-order tilt-to-length (TTL) coupling. 

\subsection{Optical setup}
\begin{figure}[ht!]    \centering\includegraphics[width=0.9\textwidth]{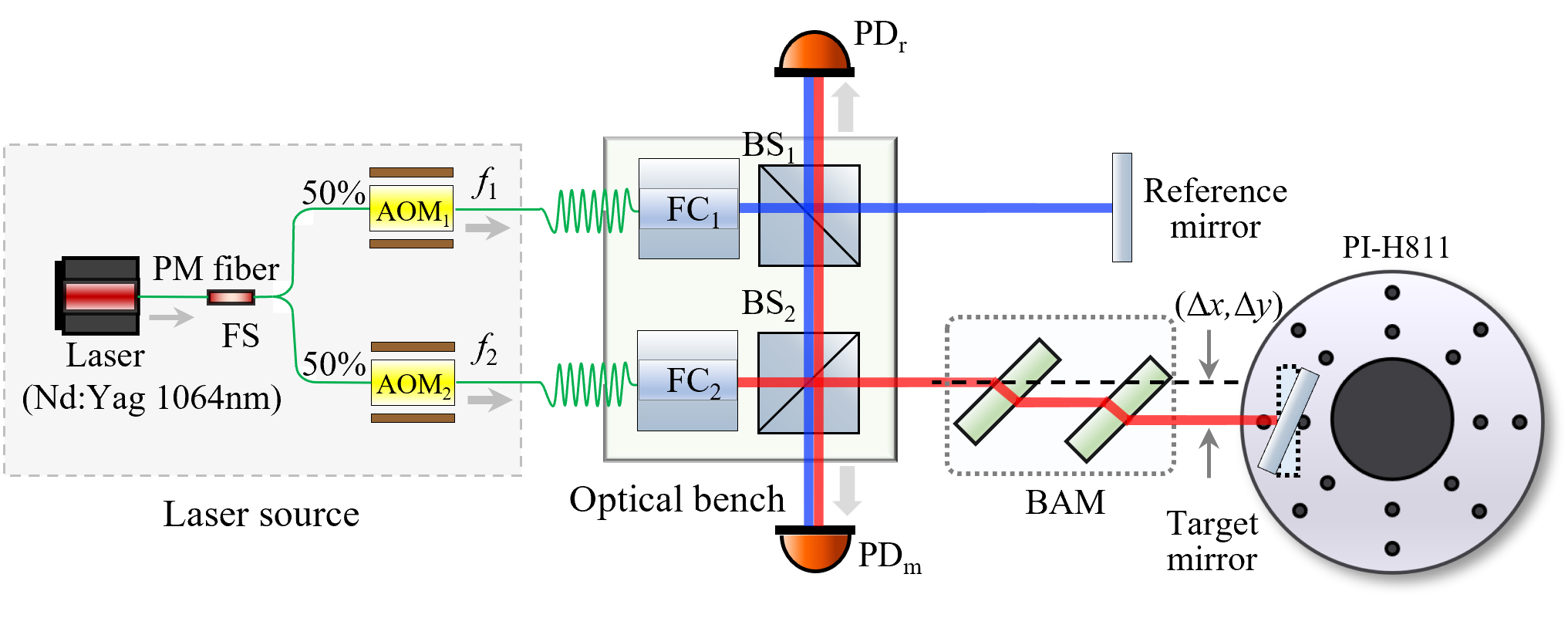}
    \caption{The optical schematic of TTL coupling suppression using the BAM. The setup incorporates critical components: FS (fiber splitter), AOM (acousto-optic modulators), FC (fiber collimator), BS (beam splitter), and PD (photodetector). The BAM introduces controlled lateral displacement to align the beam with the target mirror, effectively minimizing first-order TTL coupling errors in the interferometric measurement.}
    \label{fig13}
\end{figure}

The experimental setup is based on a compact heterodyne interferometer, designed for high-precision measurement and simplified system architecture \cite{yan_highly_2020}. The interferometer is assembled using UV bonding technology, which enhances mechanical stability and ensures long-term alignment precision \cite{lin_compact_2024-1,lin_construction_2023}. The optical \D{schemetic} \R{schematic} of TTL coupling suppressing by BAM is illustrated in \fref{fig13}. The system utilizes two heterodyne beams, differing by \SI{10}{\kilo\hertz} in frequency. These beams are introduced through collimators and directed through distinct optical paths. 

One beam is reflected off a stable reference mirror before being combined with the second beam at beam splitter $\mathrm{BS_1}$, producing a reference signal that is detected by photodetector $\mathrm{PD_r}$. The second beam traverses the BAM, undergoes reflection from a target mirror, and retraces its path back through the BAM. Finally, the beams recombine at beam splitter $\mathrm{BS_2}$, with the resulting interference signal captured by photodetector $\mathrm{PD_m}$.

The primary factor influencing first-order TTL coupling is the lateral displacement between the beam’s reflection point on the target mirror and the mirror’s rotation axis. A 6-axis Hexapod (PI-H811) is employed to control and fine-tune the target mirror’s rotation axis \R{while actively inducing sinusoidal angular jitter with a amplitude of \SI{100}{\micro\meter} and a frequency of \SI{1}{Hz}}. The BAM allows precise adjustments to the beam position on the target mirror, effectively compensating for first-order TTL coupling errors and enhancing the overall measurement accuracy.

\subsection{Test results}
\begin{figure}[ht!]
    \centering
    \includegraphics[width=0.55\textwidth]{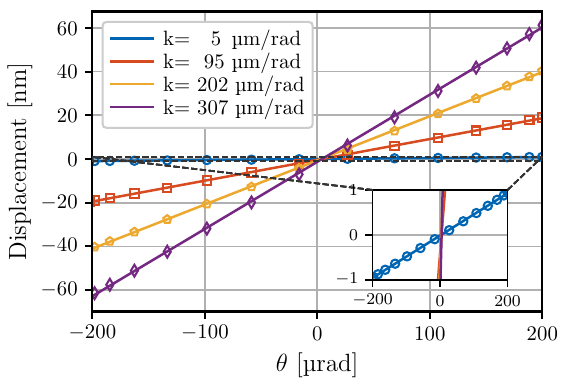}
    \caption{TTL coupling test results, showing the TTL \D{couping} \R{coupling} curve with angle. The \D{maker} \R{marker} represents the experimental data, and the line represents the fitting curve.}
    \label{fig14} 
\end{figure}

The results of the TTL tests are illustrated in Figure \ref{fig14}, which clearly shows the effectiveness of the BAM in reducing first-order TTL coupling errors. Initially, the TTL coupling coefficient was measured at approximately \SI{300}{\micro\meter/\radian}. As the BAM was employed to adjust and align the beam axis, a gradual decrease in this coupling error was observed. By meticulously fine-tuning the lateral displacement of the beam axis, the coupling coefficient was ultimately reduced to an impressive \SI{5}{\micro\meter/\radian}. Further first-order TTL coupling suppression is limited by the optical axis tuning accuracy of the BAM and the displacement measurement noise floor of the interferometer bench.

This significant reduction underscores the BAM’s potential to enhance measurement precision in interferometric systems, particularly in the context of gravitational wave detection. The experimental data aligns well with theoretical predictions, demonstrating that the BAM can effectively manage the adverse effects of misalignment and surface errors. Such capabilities are crucial for improving the performance of gravitational wave detectors, where minimizing noise is essential for accurate measurements.
This significant reduction underscores the BAM’s potential to enhance measurement precision in interferometric systems, particularly in the context of gravitational wave detection. The experimental data aligns well with theoretical predictions, demonstrating that the BAM can effectively manage the adverse \D{effects of misalignment and surface errors} \R{effects of first order TTL induced through optical component misalignments}. Such capabilities are crucial for improving the performance of gravitational wave detectors, where minimizing noise is essential for accurate measurements.

\section{Discussion and conclusion}
\label{sec_conclusion}
This study explored the Beam Alignment Mechanism's (BAM) ability to mitigate tilt-to-length (TTL) coupling noise in gravitational wave detectors. The experimental results affirm that BAM significantly compensates for lateral displacement errors, effectively aligning the optical axis and reducing TTL noise. However, challenges remain due to misalignments that can introduce additional tilts, potentially affecting suppression efficiency. While current configurations show promise, further optimization is necessary.

In conclusion, BAM represents a substantial advancement in addressing TTL noise, enhancing the precision of interferometric measurements. Ongoing research into alignment techniques and adaptive control algorithms will be vital in refining BAM’s effectiveness. By overcoming existing limitations, BAM can play a pivotal role in the space generation of gravitational wave detectors.

\section*{Data availability statement}
The data cannot be made publicly available upon publication because no suitable repository
exists for hosting data in this field of study. The data that support the findings of this study are
available upon reasonable request from the authors.

\ack
We gratefully thank Prof. Jun Luo for the helpful suggestion. This work was supported by the National Natural Science Foundation of China (Grant Nos. 12105375) and the National Key Research and Development Program of China (Grant Nos. 2023YFC2205800).

\section*{Reference}
\bibliography{bam}

\appendix
\section{Error analysis}
\subsection{Axis misalignment}
\label{app_axis}
As shown in \Fref{fig4}, point $O$ serves as both the point of incidence and the origin of the coordinate system. The unit vectors $\vec{x}$, $\vec{y}$, and $\vec{z}$ represent the axes of this coordinate system. The rotation axis is defined by its direction ($\vec{n}_{\rm{axis}}$) and a point $A$ on it, such as:
\begin{equation}
    \cases{
    A = (x_A, y_A, z_A)\\
    \vec{n}_{\rm{axis}}= (n_x, n_y,n_z)^{\rm{T}},}
\end{equation}
which satisfies the conditions:
\begin{equation}
    \cases{|\vec{n}_{\rm{axis}}| = 1\\
    \frac{\vec{n}_{\rm{axis}} \cdot \vec{y}}{|\vec{n}_{\rm{axis}}||\vec{y}|} = \cos {\delta \theta}.\\}
\end{equation}

Follow the Rodrigues' rotation formula, the rotation matrix $\mathbf{R}(\vec{n}_{\text{axis}}, \varphi)$ for an angle $\varphi$ around the rotation axis is:
\begin{equation}
    \mathbf{R} = \mathbf{I} + \sin \varphi \mathbf{K} + (1 - \cos \varphi) \mathbf{K}^2,
    \label{eq_rod}
\end{equation}
where $\mathbf{I}$ is the identity matrix and $\mathbf{K}$ is the cross-product matrix of $\vec{n}_{\text{axis}}$:
\begin{equation}
    \mathbf{K} =\left(
    \begin{array}{ccc}
    0 & -n_z & n_y \\
    n_z & 0 & -n_x \\
    -n_y & n_x & 0
    \end{array}\right).
    \label{eq_k}
\end{equation}
Initially, the normal vector of the plate surface lies in the $y\text{-}O\text{-}z$ plane as
\begin{equation}
    \vec{n}_{\rm plate} = (0, \cos \theta_0, \sin \theta_0)^{\rm T}.
\end{equation}
After rotation, the new normal vector is given by:
\begin{equation}
    \vec{n'}_{\rm plate}
    = \left(
    \begin{array}{c}
    x \\
    y \\
    z
    \end{array}\right)+
    \mathbf{R}(\vec{n}_{\text{axis}}, \varphi) \cdot  
    \left(\vec{n}_{\rm plate}-
    \left(
    \begin{array}{c}
    x \\
    y \\
    z
    \end{array}\right)\right)
    = (n'_x, n'_y,n'_z)^{\rm{T}},
    \label{eq_n}
 \end{equation}
where
\begin{equation}
    \fl\cases{
        n'_x = (1 - \cos \varphi) n_x (\cos \theta_0 n_y + \sin \theta_0 n_z) + \sin \varphi (\sin \theta_0 n_y - \cos \theta_0 n_z) \\
        n'_y = \cos \theta_0 \left(1 - (1 - \cos \varphi)(n_x^2 + n_z^2)\right) + \sin \theta_0 ((1 - \cos \varphi)n_y n_z - \sin \varphi n_x) \\
        n'_z = \cos \theta_0 ((1 - \cos \varphi)n_y n_z + \sin \varphi n_x) + \sin \theta_0 (1 - (1 - \cos \varphi)(n_x^2 + n_y^2))
    }
\end{equation}

Replacing $\vec{n}_{\rm plate}$ with $\vec{n'}_{\rm plate}$, the incident angle $\theta$ can be expressed as a function of the rotation angle $\varphi$:
\begin{equation}
    \theta(\varphi) = \arccos  \left(  \frac{\vec{n'}_{\rm plate} \cdot \vec{y}}{|\vec{n'}_{\rm plate}||\vec{y}|} \right) = \arccos n'_y
\end{equation}

\subsection{Surface parallelism error}
\label{app_surf}
As shown in \Fref{fig8}, the normal vector of the front surface is defined as:
\begin{equation}
    \vec{n}_{\rm plateA} = (0, \cos \theta, \sin \theta)^{\rm T}.
\end{equation}

The normal vector of the second surface is given by:
\begin{eqnarray}
    \vec{n}_{\rm plateB} = \mathbf{R}(\vec{n}_{\rm plateA}, \phi) \cdot (0, \cos (\theta - \alpha), \sin (\theta - \alpha))^{\rm T}\\
    = \left(\begin{array}{c}
        -\sin \phi \sin \theta \cos \theta_1 + \sin \phi \cos \theta \sin \theta_1 \\
        (1 - (1 - \cos \phi) \sin^2 \theta) \cos \theta_1 + (1 -\cos \phi) \sin \theta \cos \theta \sin \theta_1 \\
        (1 - \cos \phi) \sin \theta \cos \theta \cos \theta_1 + (1 - (1 - \cos \phi) \cos^2  \theta) \sin \theta_1
    \end{array}\right),
\end{eqnarray}
where $\theta_1 = \theta - \alpha$ and $\phi$ is the angle between the projected vector of $\vec{n}_{\rm plateA}$ in the $x\text{-}O\text{-}z$ plane and the $z$ axis. 

According to the equation (\ref{eq_n}), the two surfaces normal vector $\vec{n'}_{\rm plate0}$ after rotation around the rotation axis with plate can be expressed as:
\begin{equation}
    \vec{n'}_{\rm plateA} = \mathbf{R}(\vec{y}, \varphi) \cdot \vec{n}_{\rm plateA}
    = \left(
    \begin{array}{c}
        \sin \varphi \sin \theta \\
        \cos \theta \\
        \cos \varphi \sin \theta
    \end{array}\right),
\end{equation}
\begin{equation}
    \vec{n'}_{\rm plateB} = \mathbf{R}(\vec{y}, \varphi) \cdot \vec{n}_{\rm plateB}
\end{equation}
The direction of the exiting beam after refraction can be calculated using Snell's law, leading to the compensation value and angle:
\begin{equation}
    \vec{b}_\text{out} = \frac{\vec{b}_\text{in}}{n} + \left(\frac{\cos \theta }{n}  - \sqrt{1 - \frac{1 - \cos^2 \theta}{n^2}}\right) \vec{n}_{\rm plate} = \mathbf{M}(\vec{n}_{\rm plate}) \cdot \vec{b}_\text{in},
\end{equation}
where $\vec{b}_\text{in}$ and $\vec{b}_\text{out}$ denote the incident and refracted beam vectors at a single interface, respectively, while $\mathbf{M}(\vec{n}_{\rm plate})$ characterizes the directional transformation matrix induced by plate surface refraction. After the beam undergoes refraction twice—once at the front and once at the rear surface of the plate—the direction vector of the exiting light can be determined. The compensation value and angle are calculated similarly to the equation (\ref{eq_dh}) and (\ref{eq_varphi}) as
\begin{equation}
    \cases{
    h' = d \sin \theta \left(1 - \frac{\cos \theta_0}{\sqrt{{n}^2 - \sin^2 \theta_0}} \right)\\
    \varphi' = \varphi\\
    \vec{b}_\text{exit} = \mathbf{M}(\vec{n'}_{\rm plateB}) \cdot \mathbf{M}(\vec{n'}_{\rm plateA}) \cdot \vec{y}
    } ,
\end{equation}
where $b_{exit}$ is the direction vector of the exiting beam after refraction at both surfaces of the plate. 

\end{document}